# THE OBSERVED DISTRIBUTION FUNCTION OF
# PECULIAR VELOCITIES OF GALAXIES

Somak Raychaudhury

*Inter-University Centre for Astronomy and Astrophysics, Pune 411 007, India;*
*Harvard-Smithsonian Center for Astrophysics, Cambridge MA;*
*and Institute of Astronomy, Madingley Road, Cambridge CB3 0HA, UK[1]*

and

William C. Saslaw

*Department of Astronomy, University of Virginia, Charlottesville, VA;*
*National Radio Astronomy Observatory[2], Charlottesville, VA;*
*and Institute of Astronomy, Madingley Road, Cambridge CB3 0HA, UK*

astro-ph/9602001   1 Feb 96

[1]  *E-mail: somak@iucaa.ernet.in*
[2]  *Operated by Associated Universities, Inc., under cooperative agreement with the National Science Foundation.*

ABSTRACT

We give the first determination of the observed peculiar velocity distribution function for a representative sample of galaxies which includes a wide range of clustering properties. We explore in detail the effects of uncertainties in sampling and in distance measures on the estimated distribution function. The observed distribution function is consistent with an earlier prediction of gravitational clustering, over the entire range of peculiar velocities, from field galaxies to rich clusters, on scales up to $50h_{100}^{-1}$ Mpc. In the simplest consistent model, most of the inhomogeneous mass of the Universe is in galaxies or their halos.

We estimate the "Mach Number" for the bulk flow within $50h_{100}^{-1}$ Mpc from us to be $\mathcal{M} = \Delta v_r / \langle v_r^2 \rangle^{\frac{1}{2}} = 599/717 \simeq 0.8$, which includes the effect of high-dispersion galaxies in clusters. The observed velocity distribution function agrees quantitatively with N-body simulations with $\Omega_0 = 1$. Further comparisons of the observed distribution with N-body simulations will provide a new technique for measuring $H_0$ and $\Omega_0$. These results provide new tests for all models of galaxy clustering.

Subject headings: cosmology: observations – galaxies: clustering – galaxies: formation – galaxies: distances & redshifts – cosmology: large scale structure of the Universe

## 1. INTRODUCTION

The velocity distribution function $f(v)\,dv$, defined as the probability of finding a galaxy with peculiar velocity (relative to the Hubble flow) between $v$ and $v + dv$, is a fundamental astronomical property of our universe which has not been measured previously for a fair sample of galaxy peculiar velocities. It is particularly important for understanding the clustering processes that have resulted in the observed non-uniform spatial distribution of galaxies. In the case of a perfect gas, $f(v)\,dv$ would be a Maxwell-Boltzmann distribution. Observed departures from the Maxwell-Boltzmann form of the velocity distribution provide insights into its origin.

Only in the last few years have secondary distance indicators ($D_n$-$\sigma$, Tully-Fisher) yielded distances to galaxies with uncertainties $\lesssim 20\%$, thus enabling peculiar velocities to be measured beyond the local Supercluster. Studies of the distribution function of peculiar velocities require a large representative sample of galaxies without any *a priori* selection related to clustering. In this respect, for example, the Aaronson *et al.* (1986) sample of peculiar velocities of spirals belonging to rich clusters would give biased estimates of $f(v)\,dv$. However, a sample of peculiar velocities of

>1300 spirals chosen without regard to their clustering properties (Mathewson *et al.* 1992) appears to be a promising database for a first analysis of the distribution of radial peculiar velocities $f(v_r)\,dv_r$.

A theory of gravitational galaxy clustering, in which most of the dynamically important mass in the Universe is assumed to be in galaxies or their halos, makes an explicit prediction (Saslaw *et al.* 1990) for the distribution function $f(v_r)\,dv_r$. This is in a form which can readily be compared with the observed distribution.

In Section 2 we describe the data used in our analyses. Section 3 gives results for $f(v_r)\,dv_r$ and compares them with a Maxwell-Boltzmann distribution and with the prediction of gravitational clustering. Section 4 discusses the effects of uncertainties in sample selection or distance measurement on our results, and in Section 5, we look at some further cosmological implications of our results.

## 2. THE SAMPLES OF OBSERVED PECULIAR VELOCITIES

### 2.1 THE MATHEWSON ET AL. CATALOG OF SPIRAL GALAXIES

The Mathewson *et al.* (1992) sample consists of peculiar velocities of 1353 spirals with redshift <7000 km s$^{-1}$. It covers about one-fourth of the sky, the distances to the spirals being measured using the $I$-band Tully-Fisher relation. All late spirals (Sb–Sd) in the ESO catalog with major diameter > 1.7 arcmin, inclination > 40° and Galactic latitude $|b| > 11°$ were included in their list. Their list, however, also included some galaxies from the observations of Haynes and Giovanelli, plus a "sprinkling" of spirals beyond 7000 km s$^{-1}$, many of which have smaller diameters and lower inclinations.

To extract a well-defined sample from the Mathewson *et al.* list, we chose only the galaxies with major diameter $\geq 1$ arcmin and disk inclination $I \geq 35°$ (so that the inclination corrections to magnitudes and rotational velocities are not too high). On the sky, we spatially restricted the sample to the area of the sky bounded by $(240° < l \leq 330°,\ 11° < b \leq 45°)$, $(330° \leq l \leq 350°,$ $30° \leq b \leq 45°)$ , $(350° \leq l \leq 35°,\ b < 30°)$ , and $(210° \leq l < 350°,\ b < -11°)$, since the sample appears less complete elsewhere. We corrected the total $I$-band magnitudes for internal extinction according to the prescription of Pierce and Tully (1988), and also for Galactic extinction (Burstein and Heiles 1982), using $A_I = 0.44A_B$.

We calculate distances to galaxies in our sample using the Tully-Fisher relation (1) below, correcting for a uniform Malmquist bias[1] corresponding to $\sigma = 0.36$, which means relative distances

---

[1] The effect of inhomogeneous Malmquist bias on the measured distances is discussed in Appendix

for single galaxies are accurate to about 18%. In addition, we assign the spirals within $2.5h_{75}^{-1}$ Mpc of the center of the clusters Fornax, A1060, Antlia, Cen30 and Cen45 to the mean distance of the cluster. For these galaxies (only 94 out of our sample of 825), the corresponding Malmquist bias used is $\frac{1}{N}$ times that for a single galaxy, where $N$ is the number of spirals in our sample belonging to that cluster (Lynden-Bell *et al.* 1988).

In the rest of the paper, wherever we quote distances in units of km s$^{-1}$, or calculate peculiar velocities, we will use $H_0 = 75$ km s$^{-1}$ Mpc$^{-1}$, which is consistent with the distance to the Fornax cluster we use to calibrate our TF relation. Radial velocities are measured, unless otherwise stated, in the CMB frame.

### 2.2 Calibration of the Tully-Fisher Relation

Both Willick *et al.* (1995), who have published preliminary results from their "Mark III" catalog of peculiar velocities, and Mathewson *et al.* (1992) calibrate their Tully-Fisher relation on several clusters of galaxies, assuming that the spirals within some angular distance of the center of each cluster are at the distance of the respective cluster. However, the work of Bernstein *et al.* (1994) and Raychaudhury *et al.* (1995) shows that in several clusters, including Coma and A2634, which Willick *et al.* use to calibrate their TF relation, most of the galaxies bound to the core of the cluster are too gas-poor to allow Tully-Fisher calibration. Here, in order to decide which clusters to use for calibration, we perform a simple test.

---

**Table 1: Spirals within $2.5h_{75}^{-1}$ Mpc of the centers of clusters**

| Cluster | Number of galaxies | Dispersion about TF relation if galaxies at | |
|---|---|---|---|
| | | same distance | redshift distance |
| Fornax | 10 | 0.23 | 0.54 |
| Hydra | 16 | 0.26 | 0.26 |
| Antlia | 20 | 0.43 | 0.37 |
| Sculptor | 20 | 0.75 | 0.46 |
| Eridanus | 11 | 0.78 | 0.78 |
| Pegasus | 11 | 0.41 | 0.30 |

---

For each of the major clusters in the Mathewson *et al.* list, we choose all galaxies that lie within $2.5\,h_{75}^{-1}$ Mpc of its center on the sky (calculated at the mean redshift of the cluster), and have a redshift within $\pm 1000$ km/s of the mean redshift. We eliminated a few galaxies from this list on inspecting their rotation curves because they were found to be rising at both ends. For each cluster, we calculate the dispersion of the magnitudes about the best-fit Tully-Fisher relation, assuming all galaxies are at the same distance. We then repeat the exercise for the same galaxies but assuming that each galaxy lies at a distance proportional to its redshift. These values are shown in Table 1. In order to calibrate the TF relation on spirals in a cluster, it is necessary to assume they are at the same distance. This assumption might not be true for the clusters Sculptor, Eridanus or Pegasus, for which the dispersion if distances are proportional to redshift is considerably lower than if the galaxies are assumed to be at their cluster centers.

Therefore our calibration uses only the three spiral-rich clusters that satisfy the criterion that the value of the dispersion in column 3 of Table 1 is not significantly higher than that in column 4, and is less than 0.5 mag. We use 46 galaxies in three clusters (Fornax, A1060=Hydra I and Antlia) to calibrate the Tully-Fisher relation that we will use in this paper to calculate distances. To the samples of 10 galaxies in Fornax, 16 in A1060 and 20 in Antlia, we fit Tully-Fisher relations (Figure 1) of the form $s = \alpha I + \beta$, where $s = \log \Delta v$ is corrected for inclination, and $I$ is the apparent $I$-magnitude corrected for extinction effects. The dotted lines represent the best-fit TF relations to each data set, where both $\alpha$ and $\beta$ were allowed to vary. However, we preferred to force $\alpha$ to be the same in all three cases, and used linear regression on $s$ to find one value of $\alpha$ and a different $\beta$ for each cluster, the latter reflecting their distance.

McMillan *et al.* (1993) have found the distance modulus to the Fornax cluster from planetary nebulae to be $\mu = 31.14 \pm 0.14$. We use this to calibrate the zero-point, which, together with the slope as obtained above, gives the Tully-Fisher relation,

$$s = -0.121\,M_I - 0.43 \tag{1}$$

for the Fornax cluster, where $M_I$ is the absolute magnitude of a galaxy in the $I$-band.

The value for the scatter in magnitude of the TF relation is obtained using

$$\sigma = \sqrt{\frac{\sum_{i=1}^{N}(M_{I,i} - M'_{I,i})^2}{N-4}},$$

where $M'_{I,i}$ is the expected value of the magnitude of the galaxy from (1). From all $N = 46$ galaxies, we obtain $\sigma = 0.36$ mag. This is not very different from the $\sigma = 0.32$ mag obtained by Mathewson *et al.*

## 2.3 The "Seven Samurai" catalog of Elliptical galaxies

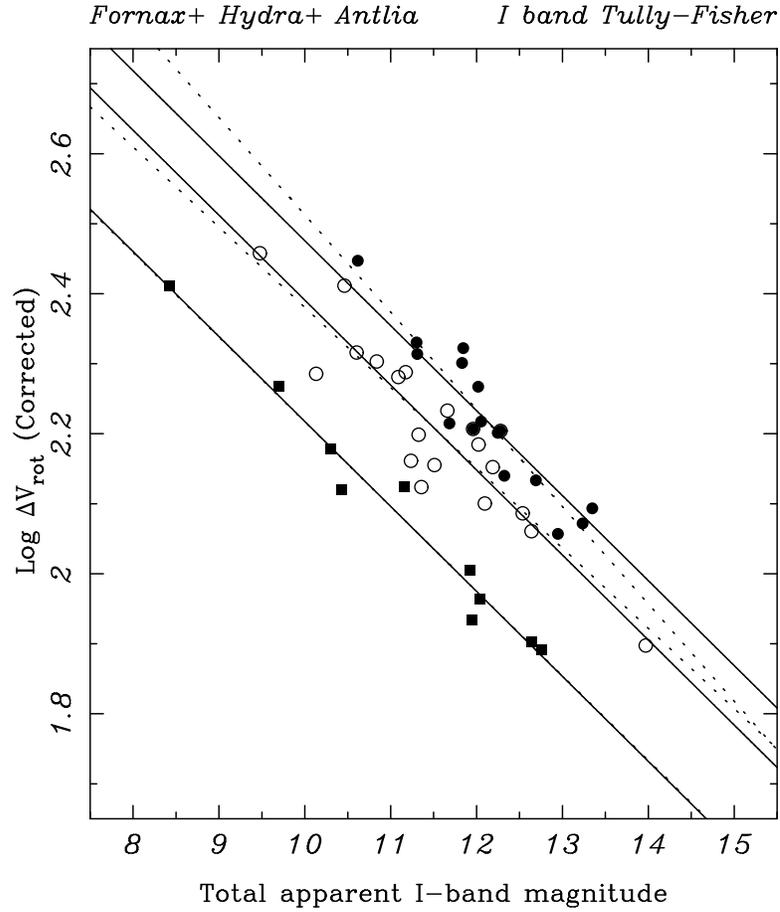

*Fornax+ Hydra+ Antlia        I band Tully-Fisher*

**Figure 1:** The *I*-band Tully-Fisher relation for galaxies belonging to three clusters of galaxies (Squares: Fornax; Filled Circles: Hydra I (A1060) and Open Circles: Antlia) in the Mathewson *et al.* (1992) catalog. All galaxies belonging to the same cluster are assumed to be at the same distance. The dotted lines are the best-fit Tully-Fisher relations for the individual clusters. The solid lines result if all clusters are required to give the same slope.

The "Seven Samurai" sample (Burstein *et al.* 1987, Dressler *et al.* 1987, Lynden-Bell *et al.* 1988) of galaxies consists of 449 elliptical galaxies chosen from all over the sky, for which distances were measured using the $D_n - \sigma$ relation. The dispersion in the logarithm of measured distance was $\Delta = 0.21$. In this paper, we also calculate $f(v_r)$ for this set of data, using a subsample consisting of the 376 elliptical galaxies with $D < 5000$ km/s, chosen from Burstein's "Mark II" catalogue (privately circulated), based mainly on the Seven Samurai sample, with a few ellipticals included in addition. We will however base most of our analyses in this paper on the Mathewson *et al.* sample because spirals are more representatively distributed, whereas the Seven Samurai sample of ellipticals might preferentially sample high density environments.

## 3. THE RADIAL VELOCITY DISTRIBUTION FUNCTION:
## OBSERVATIONS COMPARED WITH THEORY

It is now well-established that galaxies within a radius of $5000 \text{ km s}^{-1}$ around us participate in a bulk flow across the whole volume over and above the Hubble expansion (Lynden-Bell *et al.* 1988, Courteau *et al.* 1993). This is presumably caused by gravitational effects of overdense regions outside this "local" volume. Since the physical model of $f(v)\,dv$ which we compare with the observations does not include bulk flows, we must first remove the bulk flow from the Mathewson *et al.* sample. Since this sample does not cover the whole sky, the mean value of its peculiar velocities would not give a reliable measure of the bulk flow.

The most accurate method presently available for this is to subtract the mean bulk motion of $599 \pm 104 \text{ km s}^{-1}$ toward $l = 312° \pm 11°$, $b = 6° \pm 10°$ found in the analysis of Dressler *et al.* (1987) of the all-sky sample of 289 elliptical galaxies with $v_{\text{obs}} < 6000 \text{ km s}^{-1}$, a subset of the "Seven Samurai" catalog referred to in §2.3. This assumes that the spirals in our sample participate in the same bulk motion as a sample of ellipticals in the same volume of space, which would be the case if these motions were purely gravitational in origin. A more recent estimate of the local value of this bulk motion comes from a study of 353 Sb-Sc spirals from the UGC catalog, with distances measured using an *r*-band Tully-Fisher relation (Courteau *et al.* 1993). They obtain a bulk motion of $360 \pm 40 \text{ km s}^{-1}$ toward $l = 294°$, $b = 0°$ out to 6000 km/s. We will use both these estimates to show the effect of the bulk flow subtraction from our sample on our results.

Alternatively, we may correct for a bulk motion of the local rest frame by subtracting an average displacement velocity $rH_0$ from each galaxy, such that the average peculiar velocity for the sample $\langle v - rH_0 \rangle = 0$. This determines an effective local value for $H_0$ in the sample, and tends to increase the net velocity dispersion of the sample. We will compare the results of both corrections.

We will also compare the observed distribution function with two theoretical distributions. The first is a prediction for non-linear gravitational clustering (Saslaw *et al.* 1990, Inagaki *et al.* 1992, Itoh *et al.* 1993). For the quasi-equilibrium gravitational clustering of point-mass galaxies in the expanding Universe, the gravitational distribution of radial peculiar velocities has the form (Inagaki, Itoh & Saslaw, 1992)

$$f(v_r) = \alpha^2 \beta (1-b) \exp\left[-\alpha\beta\left(1-b\right)\right]$$
$$\times \int_0^\infty \frac{v_\perp}{\sqrt{v_r^2 + v_\perp{}^2}} \frac{[\alpha\beta\left(1-b\right) + \alpha\left(v_r^2 + v_\perp{}^2\right)b]^{\alpha\left(v_r^2 + v_\perp{}^2\right)-1}}{\Gamma\left[\alpha\left(v_r^2 + v_\perp{}^2\right)+1\right]}$$
$$\times \exp\left[-\alpha b\left(v_r^2 + v_\perp{}^2\right)\right] dv_\perp, \quad (2)$$

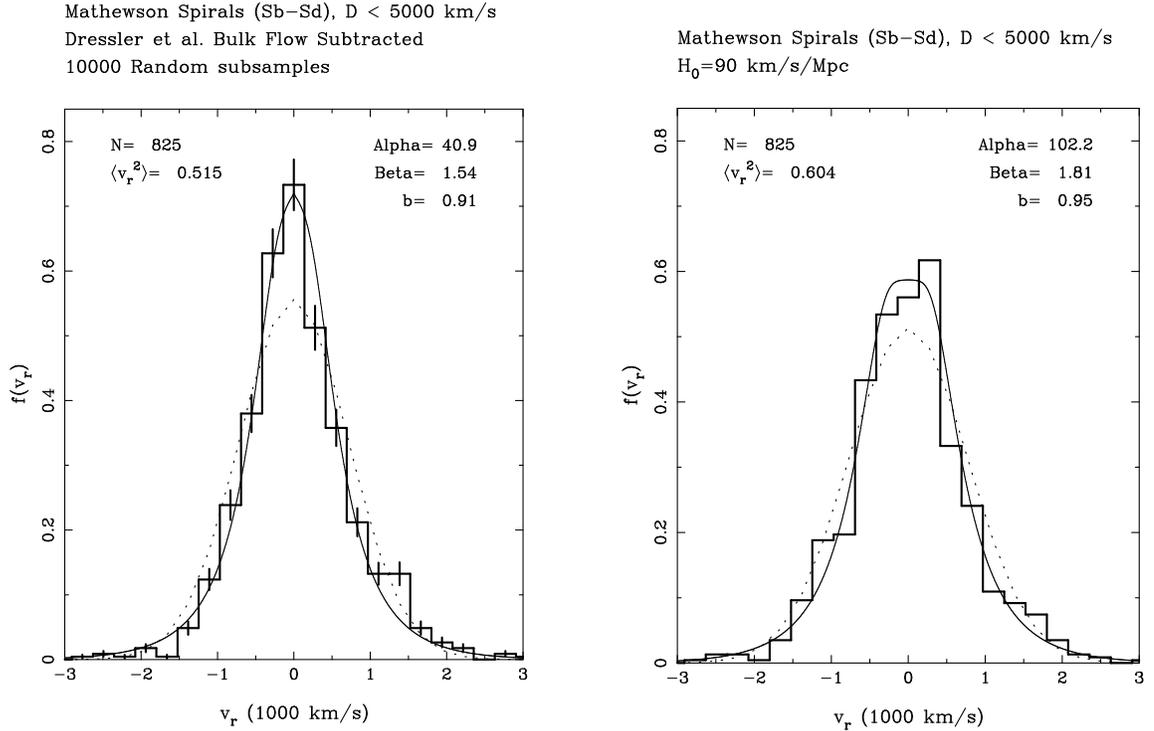

**Figure 2:** The observed distribution (solid histogram) of radial peculiar velocities (unit: 1000 km/s) for our selected subsample of the Mathewson *et al.* catalog. In (a), the peculiar velocities have been corrected for the Dressler *et al.* (1986) bulk motion of 599 km s$^{-1}$. The error bars are explained in Section 3. In (b), the peculiar velocities have been corrected for an extra Hubble expansion as discussed in Section 3. The dotted line in each case is the Maxwell-Boltzmann distribution with the same velocity dispersion as the sample. The solid line is the best-fitting gravitational distribution, whose parameters are given in the upper right-hand corner.

where $v_r$ and $v_\perp$ are the radial and transverse components of the peculiar velocity of a galaxy, and $\Gamma(x)$ is the standard Gamma Function. The quantities $\alpha$, $\beta$ and $b$ are determined by the gravitational theory, but here we shall consider them to be parameters found by fitting (2) to the observed distribution. Secondly, we will consider a Maxwell-Boltzmann distribution, since the difference between these two distributions illustrates the departure from an uncorrelated system.

Figures 2 show the velocity distributions of the sample of Mathewson *et al.* spirals in our chosen region (§2.1) with distance $D \leq 5000$ km s$^{-1}$. In Figure 2a, the detailed bulk flow obtained by Dressler *et al.* is subtracted. The histogram shown is the mean of 10000 histograms, each calculated from a randomly selected subsample consisting of two-thirds of the entire sample of 825 galaxies. The mean histogram is identical to the corresponding histogram for the entire sample with the same binning. The plotted errors are the values for the standard deviation obtained for each bin from the 10000 subsamples. We do this exercise only for this plot, to show the typical magnitude of sampling errors. The solid line is the gravitational velocity distributions (2) with the corresponding values of the best-fit parameters given in the upper right of each panel. For this figure only, we have set $\beta = 3\langle v_r^2 \rangle$; in the rest of the paper, $\beta$ will be a free parameter. Also, in most of the subsequent plots, we will only show histograms for the entire sample without error bars.

In Figure 2b, instead of the bulk flow, an extra Hubble expansion is subtracted from the observed peculiar velocities such that $\langle v_r - rH \rangle = 0$, which requires $H = 92$ km s$^{-1}$ Mpc$^{-1}$. Although this is slightly higher than the value of $H_0 = 75$ km s$^{-1}$ Mpc$^{-1}$ we use here, this represents an *effective* local value of $H_0$, and in the presence of bulk flows does not give the true value of the Hubble constant. The dotted lines in both cases are Maxwell-Boltzmann distributions with the same $\langle v_r^2 \rangle$ as the data.

It is clear that the predicted gravitational distribution gives a good fit to the data over the entire range from field galaxies to those in rich clusters (e.g., the Chi-square parameter $Q = 1.0$ for Figure 2a). The Maxwell-Boltzmann distribution, however, fails to represent the large number of low velocity galaxies (mainly field galaxies) in the peak, and systematically gives too many intermediate velocity galaxies.

We have examined $f(v_r)$ for a more clustered subset, containing 194 of these spirals, which consists of all spirals lying within a cone of radius $3h_{75}^{-1}$ Mpc of the center of each cluster given in Table 3 of Mathewson *et al.* (at the mean redshift of the cluster). Although there is some contamination from projection, more rigorous definitions of galaxies in clusters are difficult due to incomplete redshift information for *all* galaxies in the volume we explore here. The velocity distribution for this more clustered subsample is much closer to a Maxwell-Boltzmann distribution than that for the total sample shown in Figure 2.

We also examined $f(v_r)$ for the "Seven Samurai" all-sky sample (described above in §2.3) of 376 elliptical galaxies with distance $D < 5000$ km s$^{-1}$ (Figure 3), Apart from being more irregular because of its small number of galaxies, the tails of this distribution are significantly skewed towards negative velocities. These properties may be caused by the preferential tendency of ellipticals to cluster (e.g. Lahav & Saslaw 1992, Dressler 1980) and by chance fluctuations in the number, positions, and relative velocities of rich clusters in this small sample. Again there is a clear departure from the Maxwell-Boltzmann distribution.

## 4. EFFECTS OF SAMPLE SELECTION AND OTHER UNCERTAINTIES

It is clearly important to examine how systematic effects and uncertainties in the data affect the velocity distribution function. We shall discuss the sources of error here with reference to the

Seven Samurai + other Ellipticals, D < 5000 km/s

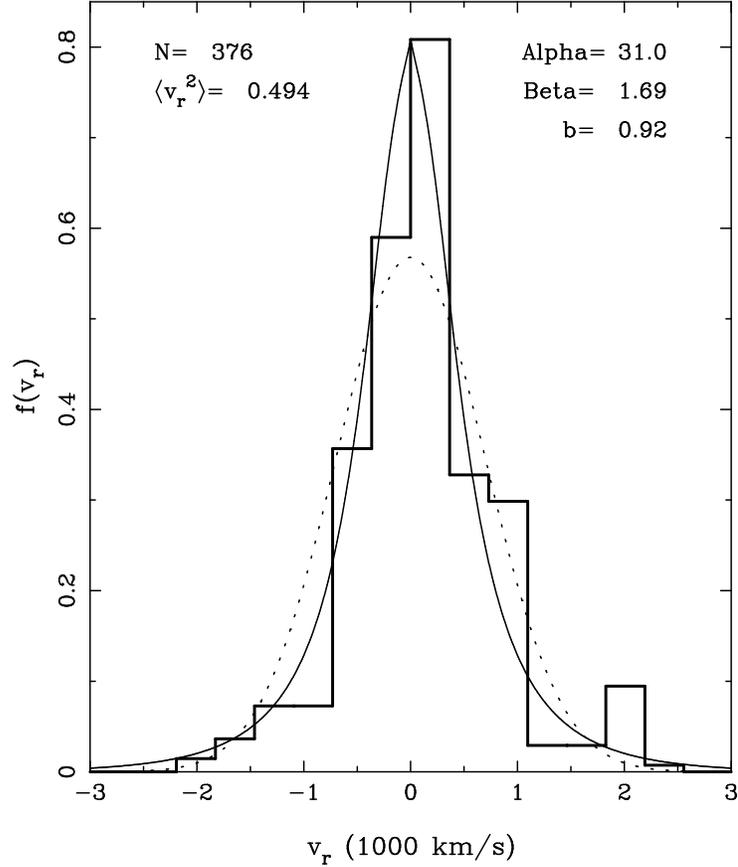

**Figure 3:** The distribution of radial peculiar velocities (unit: 1000 km/s) for all elliptical galaxies in the "Seven Samurai" sample with distance $D < 5000$ km s$^{-1}$. The dotted line is the Maxwell-Boltzmann distribution with the same velocity dispersion as the sample. The solid line is the best-fitting gravitational distribution, whose parameters are in the upper right-hand corner.

Mathewson *et al.* sample (MFB) that we use in Figure 2. We have already discussed the sampling errors in §3, which are reflected in the error bars drawn in Figure 2a.

### 4.1 LIMITING BY REDSHIFT INSTEAD OF DISTANCE

In Figures 2, we limited our sample of galaxies by measured distance, in spite of the fact our distances are more uncertain than the redshifts of the galaxies. Limiting a sample by redshift, however, introduces a systematic bias in the peculiar velocity distribution as clearly shown in Set

Mathewson Spirals (Sb–Sd), V < 5000 km/s
Dressler et al. Bulk Flow Subtracted
10000 Random subsamples

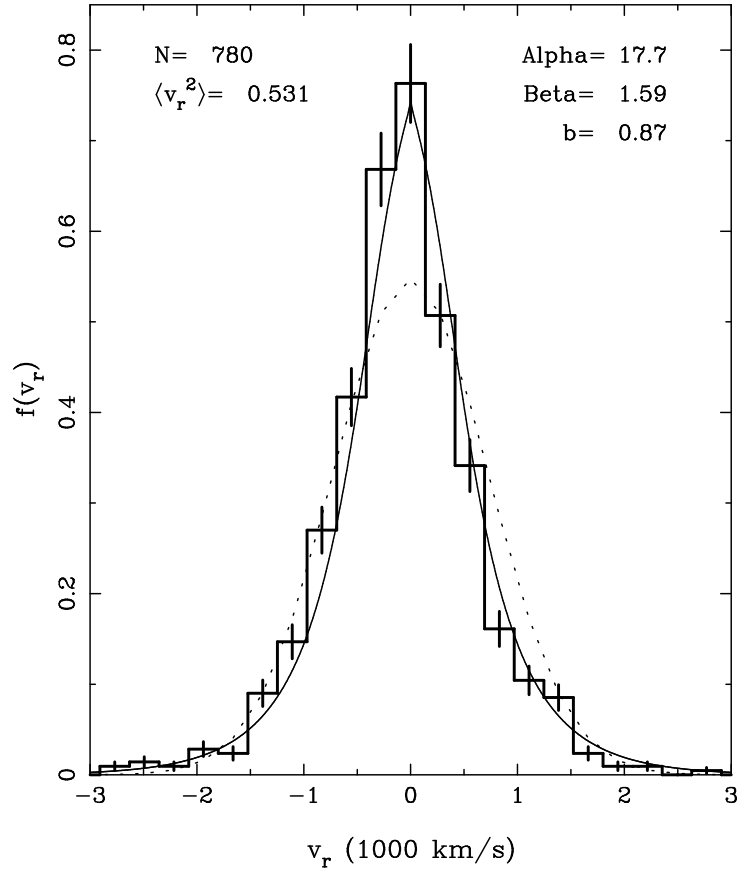

**Figure 4:** The effect of limiting the sample by redshift rather than measured distance. This plot is similar to Figure 2a, but shows all galaxies with $V < 5000$ km/s instead. The dotted line is the Maxwell-Boltzmann distribution with the same velocity dispersion as the sample. The solid line is the best-fitting gravitational distribution, whose parameters are in the upper right-hand corner.

#5 of Table 2. However, in Figure 4 we show that for the 780 galaxies satisfying the condition $v_{\mathrm{CMB}} \leq 5000$ km s$^{-1}$, the results are not substantially different from Figure 2a as far as the velocity distribution function is concerned.

### 4.2 BULK FLOW SUBTRACTION

The second uncertainty arises from the error in the subtraction of the bulk flow from the individual peculiar velocities. This is particularly important since the MFB sample does not cover

---

Table 2 goes here

---

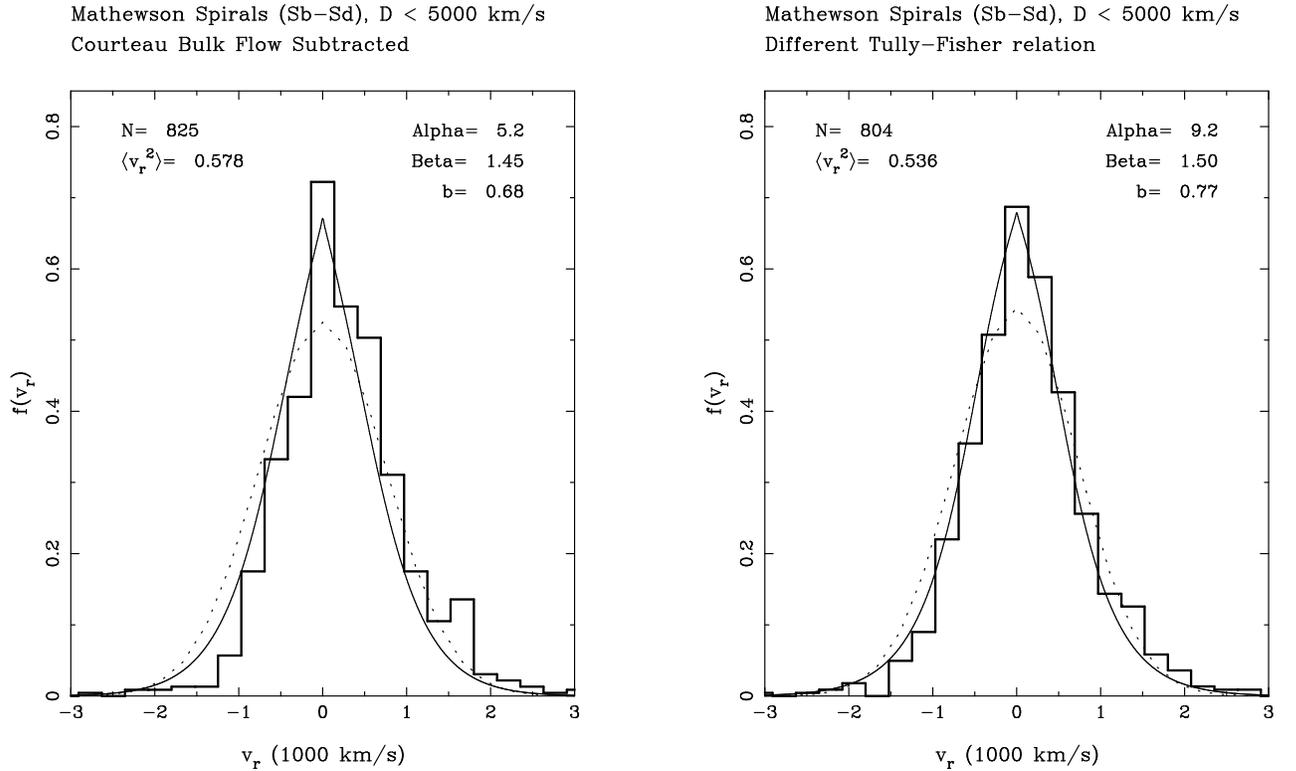

**Figure 5:** The radial velocity distribution function of the same galaxy sample as in Figure 2, but with a different bulk flow subtraction ($360 \pm 40$ km s$^{-1}$ toward $l = 294°$, $b = 0°$, Courteau *et al.* 1993). The histogram is not symmetric about $v_r = 0$, which might indicate that the subtracted bulk flow is too small in amplitude. The dotted line is the Maxwell-Boltzmann distribution with the same velocity dispersion as the sample. The solid line is the best-fitting gravitational distribution, whose parameters are in the upper right-hand corner.

**Figure 6:** The effect of the slope and scatter of the Tully-Fisher relation on the observed distribution function. With the sample of Figure 2a, we now use a Tully-Fisher relation with a shallower slope ($a = 6.7$, $I$ vs $s$), and larger dispersion (0.45 mag), obtained from Willick *et al.* (Private communication). The dotted line is the Maxwell-Boltzmann distribution with the same velocity dispersion as the sample. The solid line is the best-fitting gravitational distribution, whose parameters are in the upper right-hand corner.

the whole sky. Without this subtraction, the histogram of $f(v_r)$ is centered at $v_r \simeq 500$ km/s. If instead of the Dressler *et al.* estimate of the bulk flow that we use in Figure 2a, we use another recent estimate (Courteau *et al.* 1993), of $360 \pm 40$ km s$^{-1}$ toward $l = 294°$, $b = 0°$, we obtain Figure 5. The histogram is not symmetric about $v_r = 0$, which indicates that the subtracted bulk flow might be too small. However, the results are qualitatively similar to those in Figure 2a.

### 4.3 A DIFFERENT TULLY-FISHER RELATION

The third uncertainty arises from our fit (Equation 1) to the Tully-Fisher relation. Note that the dispersion in the TF relation we use here $\sigma = 0.36$ mag is almost the same as MFB's own estimate ($\sigma = 0.32$ mag), but considerably less than some other recent estimates ($\sigma = 0.45$ mag, Willick *et al.*, private communication; or 0.5–0.6 mag, Federspiel *et al.* 1994). Instead of choosing the whole MFB sample, we have selected a subsample that is more homogeneous, and yet not biased *a priori* for clustering, as described in §2. In this section, we investigate how different values of the slope and intercept in the TF relation affect our $f(v_r)\,dv_r$.

We have recomputed (Figure 6) $f(v_r)\,dv_r$ for a Tully-Fisher relation of the form $I = as + b$, with the slope $a = 6.7$ and dispersion 0.45 mag, which is the most recent estimate of Willick *et al.* (private communication) from a re-analysis of the Mathewson *et al.* sample. Here we follow Willick *et al.* and use the Courteau *et al.* correction for bulk flow (§4.2), as in the previous section. The increased uncertainty in distance measurements results in a decreased difference between the relevant Maxwell-Boltzmann distribution and the best fit to (2), but the latter is still a better fit.

### 4.4 RANDOM ERRORS IN PECULIAR VELOCITY

To test how robust the velocity distribution is to random peculiar velocity errors, we added random radial velocities, drawn from a Gaussian distribution with $\sigma = 400$ km/s, to each peculiar radial velocity in the homogeneous subsample of Figure 2a. The result (Figure 7) broadened the distribution from $\beta=1.5$ to $\beta=2.3$ (the value of $b$ remaining at $\sim$0.9), and increased the resemblance of the homogeneous subsample to the entire Mathewson *et al.* sample, and to a Maxwell-Boltzmann distribution. However, to affect the observational result substantially, the random errors are required to have a dispersion much more than 0.5 times the value of $\langle v_r \rangle$ for the data, which is unrealistically large.

### 4.5 RANDOM ERRORS IN DISTANCE MEASURES

The fifth uncertainty arises from random errors in the TF magnitudes for galaxies in our subsample. Often it is thought that because distances are exponentially related to magnitudes, a Gaussian distribution of magnitude errors will produce a lognormal distribution in distance (and therefore of peculiar velocity) errors. The actual situation is more complicated, however, and the lognormal distribution for distance errors departs from a Gaussian only when the fractional distance errors become large.

To see this, consider a simple model which illustrates how observational errors $\Delta M$ in the inferred absolute magnitude $M_0$ of the galaxies affect the distribution $f(v_r)$ for galaxies at a given

Mathewson Spirals (Sb–Sd), D < 5000 km/s
Gaussian Noise in Peculiar Velocity added

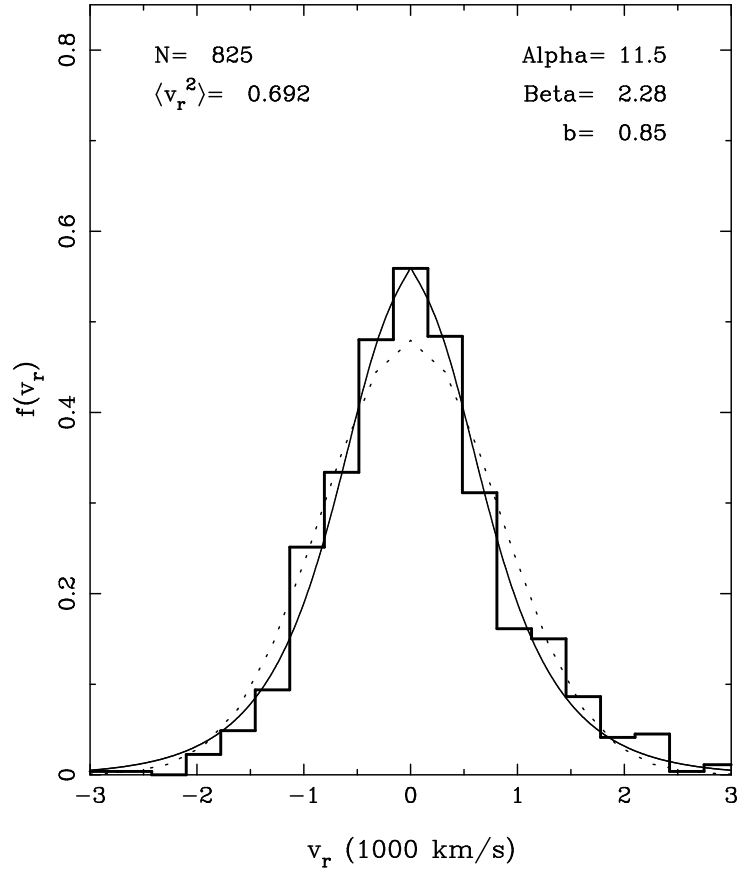

N= 825       Alpha= 11.5
$\langle v_r^2 \rangle$= 0.692       Beta= 2.28
                  b= 0.85

**Figure 7:** The effect of measurement errors in peculiar velocity on the distribution function. To each peculiar velocity in Figure 2a, we add random radial velocities drawn from a Gaussian distribution with $\sigma = 400$ km/s. This increases the resemblance of the homogeneous subsample to a Maxwell-Boltzmann distribution. In order to affect the observational result substantially, $\sigma$ for the random errors must be $\gg 0.5$ times the value of $\sigma$ for the data. The dotted line is the Maxwell-Boltzmann distribution with the same velocity dispersion as the sample. The solid line is the best-fitting gravitational distribution, whose parameters are in the upper right-hand corner. sr1 2–Aug–1995 14:02

distance $r_0$. The distance $r$ (in Mpc), which an observer will ascribe to a galaxy of apparent magnitude $m$ is

$$r = 10^{-5}.10^{0.2\,[m-(M_0+\Delta M)]}$$
$$= r_0 \exp(-a\Delta M), \tag{3}$$

Mathewson Spirals (Sb−Sd), D < 5000 km/s
Gaussian Noise in Velocity & Distance added

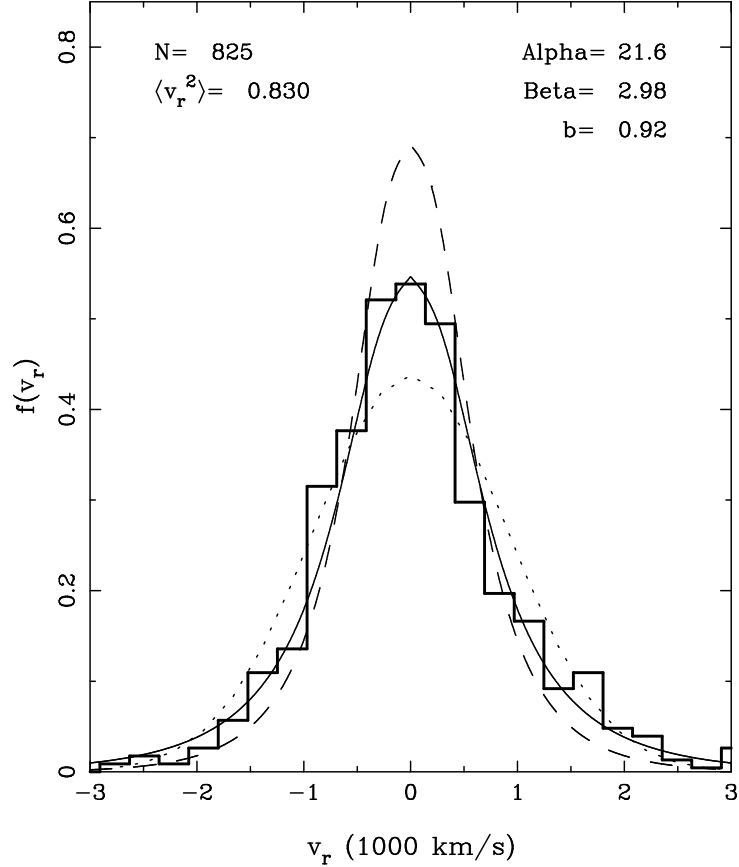

**Figure 8:** The effect of random errors in distance measurements on the peculiar velocity distribution. The measured redshifts and magnitudes of the galaxies are perturbed by random quantities as discussed in §4.5. This demonstrates that even the largest distance errors expected due to the dispersion in the Tully-Fisher relation fail to destroy the departure of the observed distribution function from a Maxwell-Boltzmann distribution. The dotted line is the Maxwell-Boltzmann distribution with the same velocity dispersion as the sample. The dashed line is the best-fit gravitational quasi-equilibrium model from Figure 2a. The solid line is the best-fitting gravitational distribution, whose parameters are in the upper right-hand corner.

where $a = 0.46$, and $\log_{10} r_0 = 0.2\,(m - M_0) - 5$. Let

$$r = r_0 \left(1 - \frac{\Delta r}{r_0}\right) \tag{4}$$

define the distance error $\Delta r$, so that using Equation (3),

$$\Delta r = r_0 \left[1 - \exp\left(-a\Delta M\right)\right]. \tag{5}$$

An observer will therefore ascribe a peculiar velocity $v_{\text{pec}}$ to the galaxy, including the error $\Delta r$, of

$$
\begin{aligned}
v_{\text{pec}} &= v_z - rH \\
&= v_z - r_0(1 - \frac{\Delta r}{r_0})H \\
&= v_z - v_{\text{Hubble}} + (\Delta r)H. \tag{6}
\end{aligned}
$$

Since $v_z$ is the measured redshift velocity (assuming no error) and $r$ is the distance where an observer believes the galaxy to be, including the effect of $\Delta M$ on $\Delta r$, an error $\Delta M > 0$ that makes the galaxy seem intrinsically fainter than it actually is will add to $v_{\text{pec}}$. If a galaxy seems intrinsically fainter than it is, $\Delta r > 0$, and from Equation (4) its actual distance $r_0$ will be greater than the ascribed distance $r$.

Now suppose the distribution of $\Delta M$ for galaxies at $r_0$ is a Gaussian of dispersion $\sigma_M^2$ centered at zero. Then from (5),

$$\Delta M = -\frac{1}{a} \ln\left(1 - \frac{\Delta r}{r_0}\right) \tag{7}$$

so that the absolute value of the Jacobian is

$$\left|\frac{\partial(\Delta M)}{\partial(\Delta r)}\right| = \frac{1}{a\left(r_0 - \Delta r\right)} \tag{8}$$

for $\Delta r < r_0$ and thus

$$f\left(\Delta r\right) = \frac{1}{\sqrt{2\pi\sigma_M^2 a^2 r_0^2}} \, \frac{1}{\left(1 - \frac{\Delta r}{r_0}\right)} \, \exp\left[-\frac{1}{2\sigma_M^2 a^2} \ln^2\left(1 - \frac{\Delta r}{r_0}\right)\right]. \tag{9}$$

This is indeed a lognormal distribution, but in $\left(1 - \frac{\Delta r}{r_0}\right)$ rather than in $\frac{\Delta r}{r_0}$ directly.

Finally, since $\Delta v_{\text{pec}} = v_{\text{pec}} - (v_z - v_{\text{Hubble}}) = H\Delta r$ from (6); and the actual Hubble velocity is $v_0 = v_{\text{Hubble}} = Hr_0$, we have

$$\frac{\Delta v_{\text{pec}}}{v_0} = \frac{\Delta r}{r_0}. \tag{10}$$

Therefore $f\left(\frac{\Delta v_{\text{pec}}}{v_0}\right)$ is lognormal in $\left(1 - \frac{\Delta v_{\text{pec}}}{v_0}\right)$.

The most probable value of $\frac{\Delta v_{\text{pec}}}{v_0} = \frac{\Delta r}{r_0} \simeq a\,\sigma_M \simeq 0.2$, using $\sigma_M = 0.4$ and $a = 0.46$. Expanding (9) for $f\left(\Delta v_{\text{pec}}\right)$ for small $\Delta v_{\text{pec}}$ gives

$$
\begin{aligned}
f\left(\Delta v_{\text{pec}}\right) = &\frac{1}{\sqrt{2\pi\sigma_M^2 a^2 v_0^2}} \left[1 + \frac{\Delta v_{\text{pec}}}{v_0} + \left(\frac{\Delta v_{\text{pec}}}{v_0}\right)^2 + \dots\right] \times \\
&\exp\left[-\frac{1}{2\sigma_M^2 a^2}\left(\frac{\Delta v_{\text{pec}}}{v_0}\right)^2 \left[1 + \frac{\Delta v_{\text{pec}}}{v_0} + \frac{11}{12}\left(\frac{\Delta v_{\text{pec}}}{v_0}\right)^2 + \dots\right]\right]. \tag{11}
\end{aligned}
$$

Therefore if the errors in the TF magnitudes for galaxies at $r_0$ produce errors in the peculiar velocities which are small compared to the Hubble velocity $r_0 H$, the distribution of these peculiar velocity errors is essentially Gaussian (Maxwell-Boltzmann Distribution).

As the peculiar velocity errors become larger, the influence of the $\frac{\Delta v_{pec}}{v_0}$ term makes the error distribution asymmetric. There does not seem to be any significant asymmetry in the observed $f(v_{pec})$. Moreover, if the true peculiar velocity distribution were Gaussian, it would convolve with a Gaussian error distribution to give an observed Gaussian peculiar velocity distribution, which is, however, not observed. This suggests that the observed departures from a Gaussian in Figures 2–3 are probably not dominated by errors in the TF magnitudes.

We can examine this further with a more detailed simulation, using the data directly. We start with the observed redshifts of the 825 galaxies in Figure 2a. To each of them, we add a velocity chosen at random from the Maxwell-Boltzmann distribution (dotted line) in Figure 2a. This gives us a more randomized sample in which to examine the effects of magnitude errors. These partially randomized velocities give new redshift distances. These new redshift distances are then perturbed by magnitude errors $\Delta M$ using Equation (3) with $\Delta M$ drawn at random from a Gaussian having $\sigma = 0.5$ mag. Subtracting these magnitude-perturbed randomized redshift distances from the originally observed redshift distances, we obtain a new set of peculiar velocities (relative to the CMB) incorporating greatly enhanced magnitude and velocity errors. Figure 8 shows the velocity distribution function for these new peculiar velocities, resulting from both Gaussian velocity perturbations and Gaussian TF magnitude perturbations. It is broader and much less strongly peaked than the originally observed $f(v_r)$, which is represented here from Fig 2a as the dashed line. The Maxwell-Boltzmann distribution that best fits the perturbed $f(v_r)$ is also shown. This demonstrates that even the largest distance errors expected due to the dispersion in the Tully-Fisher relation fail to destroy the departure of the observed distribution function from a Maxwell–Boltzmann distribution. It also suggests that the true $f(v_r)$ may be somewhat more peaked than the observed one.

### 4.6 Systematic errors in distance measures

Here we estimate the contributions of systematic errors in the measures of peculiar velocity that arise from distance measurement errors that increase with the distance of the galaxy. We would like to examine whether such errors lead distributions to be more centrally peaked than the Maxwell-Boltzmann distributions, as seen in the observed samples.

For each of the 825 galaxies used in Figure 2a, we replace its peculiar velocity with a value randomly drawn from a Gaussian distribution with $\sigma = 500$ km s$^{-1}$, which is smaller than in our observed sample. The dashed line in Figure 9 shows the distribution of peculiar velocities. We retain the original positions and distances, so that there are exactly the same number of galaxies

Mathewson Spirals (Sb–Sd), D < 5000 km/s
Gaussian peculiar velocities, distance dependent errors
10000 Random subsamples

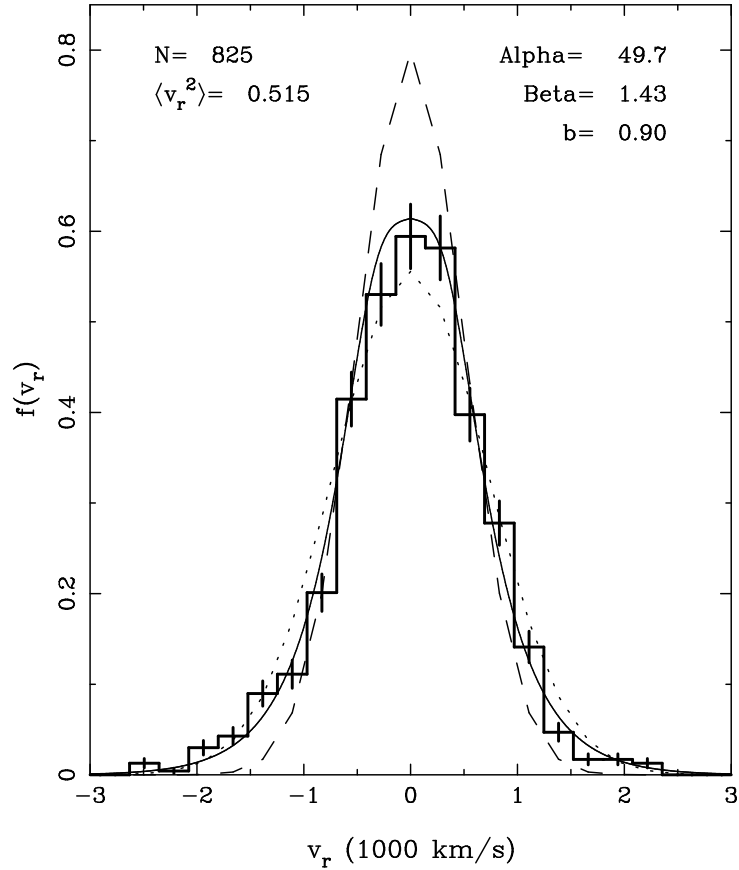

**Figure 9:** The effect of distance-dependent measurement errors on the distribution function. The dashed line is a Maxwell-Boltzmann distribution with $\sigma = 500$ km s$^{-1}$. This distribution is perturbed by magnitude errors that are proportional to the distance of each galaxy, calculated as detailed in §4.6. The dotted line is the Maxwell-Boltzmann distribution with the same velocity dispersion as the sample. The resultant histogram is less peaked in the center than the Maxwell-Boltzmann distribution we started off with, showing that such errors cannot produce the centrally-peaked observed distributions shown in the previous figures. The solid line is the best-fitting gravitational distribution, whose parameters are in the upper right-hand corner of each plot. It is very similar to the Maxwell-Boltzmann distribution with the same dispersion as the perturbed distribution.

in each of the five distance bins as shown in Table 2. Set #6 of Table 2 shows the mean peculiar velocity and rms dispersion in each distance bin: there is no systematic trend with distance.

For galaxies in each distance bin, we perturb their observed magnitudes by errors $\Delta M$ that are drawn at random from a Gaussian whose $\sigma$ is proportional to the mean distance of the bin. These magnitude errors are converted into distance errors through Equation (5), and "observed" peculiar velocities are calculated. The constant of proportionality is adjusted such that the value of $\langle v_{\text{pec}}^2 \rangle^{\frac{1}{2}}$ for the whole sample is $717$ km s$^{-1}$, the same as that is Figure 2a (compare column 5 of Sets #2 and #7 in Table 2. As the last set in Table 2 shows, this reproduces the kind of systematic increase of the value of $\langle v_{\text{pec}}^2 \rangle^{\frac{1}{2}}$ in each bin as seen in the observed sample from Set #2, Table 2). However, the resultant histogram of peculiar velocities, as seen in Figure 9, shows that the effect of such errors is quite the opposite. Starting from a narrower and peaked Maxwell-Boltzmann distribution (dashed line), the effect of the distance-dependent errors has resulted in a wider, less-peaked distribution, which shows that such errors cannot convert a sample of random peculiar velocities into a more peaked distribution as is seen in our observed samples here. Instead, the resultant distribution is still very close to a Maxwell-Boltzmann distribution (best-fit shown as dotted line).

## 5. DISCUSSION

We have determined, for the first time, the peculiar radial velocity distribution function of the galaxies. It applies to scales $\lesssim 50\, h_{100}^{-1}$ Mpc, and agrees with an earlier prediction for non-linear gravitational clustering, where most of the non-uniform dark matter in the Universe is associated with galaxies.

We use a reasonably homogeneous subsample of the Mathewson *et al.* (1992) catalog of peculiar velocities of nearby spiral galaxies, for which velocity-independent distances were measured. Crucially, these galaxies are chosen without any *a priori* bias regarding their clustering environment; this helps ensure a fair sample. To explore the effects of sampling, we have also determined $f(v_r)$ for other, less appropriate samples in addition to those discussed above. One was the entire Mathewson *et al.* sample of 1353 galaxies from which our sample of Figure 2 was chosen. We corrected each peculiar velocity for the Dressler *et al.* bulk flow correction, as in Figure 2a. The distribution function for the entire sample is broader than the homogeneous subsample, with $\langle v_r^2 \rangle^{\frac{1}{2}} = 938$ km s$^{-1}$. The fit to the gravitational distribution yields $\alpha$=11.6, $\beta$=2.57 and $b$=0.88. The increased velocity dispersion of the entire Mathewson *et al.* sample is expected from its greater heterogeneity, and the larger absolute uncertainties in peculiar velocities for galaxies beyond $D$ =5000 km s$^{-1}$.

The most straightforward interpretation of the agreement between the observed $f(v_r)$ and the theoretical prediction of (2) is that the peculiar velocities of the galaxies are caused by simple gravitational interactions. There is no obvious evidence in these data for effects of pre-galactic explosions, cosmic strings, domain walls or dynamically important dark matter not associated with galaxies or clusters. If such processes ever existed, then they would have to be compatible with these observations. On larger scales, it is not yet clear whether the bulk flow is produced by coherent initial conditions, or by chance accumulations.

Equation (2) holds under rather general conditions, which are discussed in detail in Saslaw and Hamilton (1984); Saslaw *et al.* (1990) and Saslaw and Fang (1995). These conditions which essentially include the gravitational interactions of galaxies as point masses in the expanding Universe lead to the spatial gravitational quasi-equilibrium distribution for the probability of finding N galaxies in a volume of size $V$

$$f(N, V) = \frac{\bar{N}(1-b)}{N!}[\bar{N}(1-b) + Nb]^{N-1}\exp{-[\bar{N}(1-b) + Nb]}, \qquad (12)$$

where $\bar{N} = \bar{n}V$. This provides a very good description of observed galaxy clustering in the Zwicky catalog (Crane & Saslaw 1986, Saslaw & Crane 1991), in the UGC and ESO catalogs (Lahav & Saslaw 1992), the Abell cluster catalog (Coleman & Saslaw 1990), the IRAS catalog (Sheth, Mo and Saslaw, 1994) and the Southern Sky Redshift Catalog (Fang & Zou, 1994). It also agrees very well with computer N-body simulations of galaxy clustering (Sheth and Saslaw 1995; Itoh, Inagaki & Saslaw 1993, and earlier references therein). This approach predicts a velocity distribution function $f(v)\,dv$ whose integral over the transverse velocities gives Equation (2) for $f(v_r)\,dv_r$ (Inagaki *et al.* 1992).

Other spatial distribution functions, such as the negative binomial or the compound lognormal, are almost numerically indistinguishable from (12) in the range of observed galaxy number counts (Sheth, Mo & Saslaw, 1994). They would give a result equivalent to (2), although these other spatial distribution functions do not yet have any physical basis apart from their resemblance to the gravitational quasi-equilibrium distribution. Moreover the negative binomial, for example, does not satisfy the second law of thermodynamics in the expanding Universe (Saslaw and Fang 1995).

To illustrate the accuracy with which $f(v_r)\,dv_r$ in Equation (2) agrees with the simulations, we have analyzed an experiment with 10,000 galaxies (kindly provided by M. Itoh and S. Inagaki), all of the same mass, starting from Poisson initial conditions in an $\Omega_0 = 1$ Universe (Model S of Itoh *et al.* 1993, at an expansion factor of 8). The histogram in Figure 10 shows $f(v_r)\,dv_r$ for this simulation. The dotted curve is a Maxwell-Boltzmann distribution with the same value of $\langle v_r^2 \rangle$ as the simulation. It has about the same shape relative to the simulation as the corresponding Maxwell-Boltzmann distribution in Figures 2 has to the observed galaxy distribution function. The solid curve in Figure 10 is not deliberately drawn through the histogram. It is the best fit of Equation (2) to the simulation.

Note that in Figures 3–5 & 10, the best fit value of $\beta$, which is the total velocity dispersion, is very close to $3\langle v_r^2 \rangle$. This indicates that the velocities are very nearly isotropic. For the data in

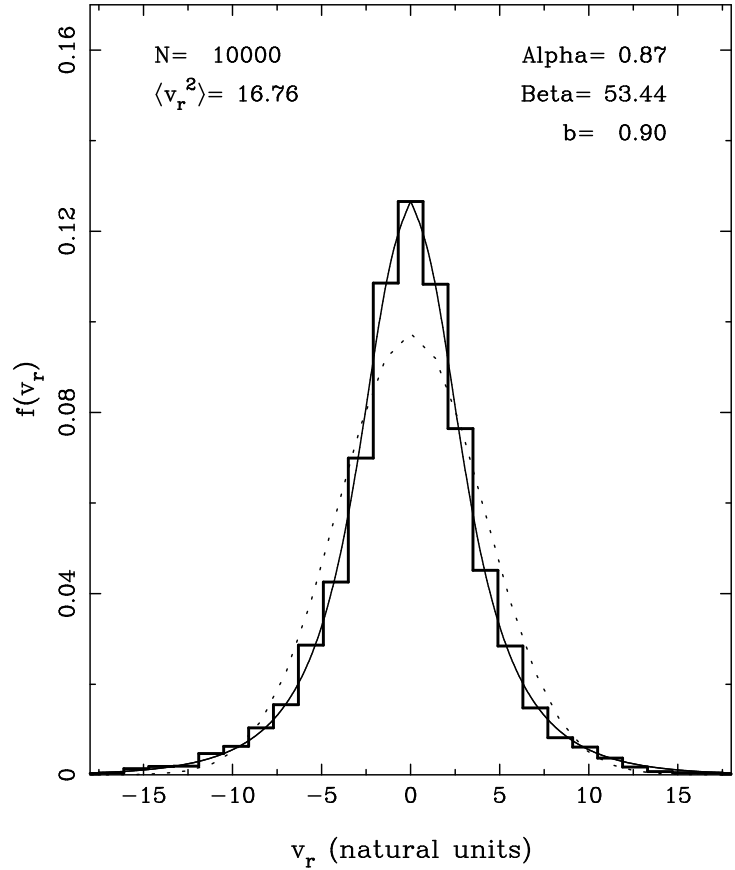

**Figure 10:** The distribution of radial peculiar velocities for the 10,000 galaxies in the simulation (model S) of Itoh **et al.** (1993), where all galaxies are of equal mass. The simulations start from Poisson initial conditions, and are observed at an expansion factor of 8. The dotted line is the Maxwell-Boltzmann distribution with the same velocity dispersion as the sample. The solid line is the best-fitting gravitational distribution, whose parameters are in the upper right-hand corner.

somak 12–Aug–1994 12:01

Figure 3, however, this relation does not hold quite so accurately, indicating some residual velocity anisotropy. Interestingly, the anisotropy appears to be larger after subtracting the Courteau *et al.* (1993) bulk flow from the velocity pattern (Figure 5) than after subtracting an extra expansion. This may be due to uncertainties in the direction and magnitude of the bulk flow, or the inappropriateness of the Courteau *et al.* flow for the region of our Mathewson subsample. We can also estimate a "Mach Number" $\mathcal{M} = \Delta v_r / \langle v_r^2 \rangle^{\frac{1}{2}} \simeq 599/717 \simeq 0.84$ for the bulk flow from Figure 2a.

This value of $\langle v_r^2 \rangle^{\frac{1}{2}}$ however includes rich clusters as well as field galaxies.

Spatial distributions of galaxies in the ESO catalog, from which most of the Mathewson *et al.* sample is drawn, are described well by the spatial distribution function of Equation (12), which is consistent with the radial velocity distribution (Lahav & Saslaw 1992). Partly as a result of sparse sampling and other selection procedures, the spatial pattern value of $b \simeq 0.6$ for large cells in the ESO may be reduced from its value of about 0.75 in the Zwicky catalog (Saslaw & Crane 1991). The numerical simulations of Figure 10 also give a spatial pattern value of $b \sim 0.75$ for large cells (Itoh *et al.* 1993). This spatial value is about 80% of $b_{\text{velocity}} \simeq 0.9$ in Figure 2a, where the Dressler *et al.* subtraction is used. Numerical simulations for a range of more realistic mass spectra than the single mass simulation in Figure 10 also generally show $b_{\text{velocity}} > b_{\text{pattern}}$, depending on the details of the mass spectrum and the value of $\Omega_0$. (Itoh *et al.* 1993). A partial reason for this difference is that in the simulations and in the observations the fits of the velocity distributions involve all the galaxies present (and all the dark matter in the observations), whereas the fits of the spatial distributions generally involve less representative subsamples.

To compare our observed distribution function with the N-body simulations of Itoh *et al.* (1993), we have to scale their velocities to the observed sample. The simulations use "natural units" with $G=m=R=1$. They are converted into physical units by (Saslaw *et al.* 1990)

$$v_{\text{scale}} = \frac{v_{\text{physical}}}{v_{\text{natural}}} = \left[ Gm \left\langle \frac{1}{r} \right\rangle \right]^{\frac{1}{2}} = \left[ 1.35 \, Gm \, N^{\frac{1}{3}} \, R^{-1} \right]^{\frac{1}{2}}, \tag{13}$$

for a simulation containing $N$ galaxies in a sphere of radius $R$, so that one natural unit of the simulation velocity equals $v_{\text{scale}}$ units of physical velocity, *e.g.* km s$^{-1}$. Assuming all of the mass $(\rho = \rho_{\text{crit}})$ in a sphere of radius $R_{\text{Mpc}}$ is associated with $N$ galaxies, the average mass of a galaxy in the simulation is $m = M/N = 6.5 \times 10^{11} N^{-1} R_{\text{Mpc}}^3 h_{75}^2 \Omega_0 (M_\odot)$, where $h_{75} = H_0/75$ km s$^{-1}$ Mpc$^{-1}$. For fixed $m$, this, together with (13) provides the velocity scale factor

$$v_{\text{scale}} = 143 \left[ \frac{N}{10^4} \right]^{-\frac{1}{3}} \left[ \frac{R}{50 \text{ Mpc}} \right] \left[ h_{75} \, \Omega^{\frac{1}{2}} \right] \text{ km s}^{-1}. \tag{14}$$

In Figure 2a, for all 825 galaxies with distance $R \leq 66\frac{2}{3}$ Mpc $(H_0 = 75)$, we find $\langle v_{r(\text{physical})}^2 \rangle^{\frac{1}{2}} = 717$ km s$^{-1}$. Continuing the comparison with the simulation in Figure 10 as an illustration, we have (model S of Itoh *et al.* 1993) $N = 10^4$, $\Omega_0 = 1$ and $\langle v_{r(\text{natural})}^2 \rangle^{\frac{1}{2}} = 4.1$. From Equation (14), this leads to a reasonable value of the Hubble parameter $h_{75} = 0.92$. However, it gives a value of $m = 1.6 \times 10^{13} (M_\odot)$, which is about an order of magnitude too high. This is a well-known result of the relatively small value of $N$ in the simulations and shows the necessity of increasing $N$ to about $10^5$ for more realistic comparisons in the $\Omega_0 = 1$ case. By increasing $N$, altering the mass spectrum, the initial conditions and the value of $\Omega_0$ in the simulations, and $H_0$ in the measured peculiar velocities, the consistency of $h$ can be improved. A systematic exploration of such models will provide a new method of determining $H_0$ and $\Omega_0$.

An understanding of the non-linear velocity distribution of galaxies has been a classic unsolved problem of cosmology ever since Milne (1935) posed it quantitatively. The agreement between our observed velocity distribution function here and the earlier non-linear gravitational prediction for this velocity distribution function is evidence that most of what we observe on these scales results from non-linear gravitational clustering. The consistency between the velocity distribution function and the spatial distribution function of galaxies is further evidence for this result. Naturally, many related questions such as the role (if any) of dark matter outside the galaxies, the effects of merging, and the detection of any memory of initial conditions still remain to be explored. However, they will all be strongly constrained to agree with this observed velocity distribution function.

## ACKNOWLEDGMENTS

We thank F. Bouchet, K. Fisher, O. Lahav, T. Padmanabhan, R. Sheth and other participants of the December 1992 workshop on *Galaxy Distribution Functions* at Pune, India, for useful comments. We thank M. Itoh and S. Inagaki for sharing their simulations with us, the referee Avishai Dekel for helpful comments, Jeff Willick for communicating his experience with the peculiar velocity data, and D. Burstein and D. Mathewson for sending us their respective catalogs of peculiar velocities in machine-readable format. We are grateful to the Tata Institute for Fundamental Research, Bombay, and the Inter-University Centre for Astronomy and Astrophysics, Pune, where part of this work was done, for their hospitality. SR has received support from a Smithsonian post-doctoral fellowship, and from NASA grant NAS8-39073 during this research. NSF grant INT-9203740 to WCS provided travel support for both of us.

## REFERENCES

Aaronson M. *et al.* 1986, ApJ, 302, 536

Bernstein, G. M., Guhathakurta, P., Raychaudhury, S., Giovanelli, R., Haynes, M. J., Herter, T. & Vogt, N. 1994, AJ, 107, 1962

Burstein, D., & Heiles, C. 1982, AJ, 87, 1165

Burstein, D., Davies, R. L., Dressler, A., Faber, S. M., Stone, R. P. S., Lynden-Bell, D., Terlevich, R., & Wegner, G., 1987, ApJS, 64, 601

Coleman, P., & Saslaw, W. C. 1990, ApJ, 353, 354

Courteau, S., Faber, S. M., Dressler, A., & Willick, J. A. 1993, ApJLett, 412, L51

Crane, P., & Saslaw, W. C. 1986, ApJ, 301, 1

Dressler, A., 1980, ApJ, 236, 351.

Dressler, A., Faber, S. M., Burstein, D., Davies, R. L., Lynden-Bell, D., Terlevich, R., & Wegner, G., *et al.* 1987, ApJ, 313, L37

Fang, F. & Zou, Z. 1994, ApJ, 421, 9.

Faber S. 1993, proceedings of the 9th IAP Meeting, Paris, on *Cosmic Velocity Fields*, ed. Bouchet, F. R. & Lachieze-Rey, M., Editions Frontières.

Federspiel, M., Sandage, A. & Tammann, G. A. 1994, ApJ, 430, 29

Inagaki, S., Itoh, M. & Saslaw, W. C. 1992, ApJ, 386, 9

Itoh, M., Inagaki, S., & Saslaw, W. C. 1993, ApJ, 403, 476

Lahav, O., & Saslaw, W. C. 1992, ApJ, 396, 430

Landy S. and Szalay A. 1992, ApJ, 391, 494

Lynden-Bell, D., Faber, S. M., Burstein, D., Davies, R. L., Dressler, A., Terlevich, R. J. & Wegner, G. 1988, ApJ, 326, 19

Mathewson, D. S., Ford, V. L., & Buchhorn, M. 1992, ApJSupp, 81, 413

McMillan, R., Ciardullo, R. & Jacoby, G. H. 1993, ApJ, 416, 62

Milne E. A. 1935, *Relativity, Gravitation and World Structure*, Oxford: Clarendon Press

Pierce, M. J., & Tully, R. B. 1988, ApJ, 330, 579

Raychaudhury, S., Bernstein, G. M. & Guhathakurta, P., 1995, *in preparation).*

Saslaw, W. C., Chitre, S. M., Itoh, M., & Inagaki, S. 1990, ApJ, 365, 419

Saslaw, W. C., & Crane, P. 1991, ApJ, 380, 315

Saslaw, W. C., & Hamilton, A. J. S. 1984, ApJ, 276, 13

Saslaw, W. C. & Fang F. 1995, to be published, ApJ

Sheth, R. K., Mo, H., & Saslaw, W. C. 1994, ApJ, 427, 562

Sheth, R. K., & Saslaw, W. C. 1995, submitted to ApJ

Strauss M. and Willick J. 1995, Phys Rep (in the press)

Tully, B. & Fisher, R. 1988, *Nearby Galaxies Catalog*, Cambridge University Press.

Willick, J. A., Courteau, S., Faber, S. M., Burstein, D., & Dekel, A. 1995, ApJ, 446, 12

## APPENDIX: THE EFFECT OF INHOMOGENEOUS MALMQUIST BIAS

If there is a dispersion in the magnitudes of galaxies, either intrinsic or from observational uncertainties, then the boundary of a magnitude-limited sample is not well-defined. There will be a tendency for more galaxies at the boundary to appear in the sample than a uniform spatial density would suggest. This is because there are more galaxies in the larger shell beyond the boundary whose magnitude dispersion makes them appear within the sample, and fewer galaxies in the smaller shell within the boundary which appear to be outside.

However this homogeneous Malmquist bias is modified by density inhomogeneities due to clustering. The resulting inhomogeneous Malmquist bias is usually a second order effect. It may be significant if there happens to be a dense cluster or large void near the boundary. These effects can modify the observed velocity distribution, and the inhomogeneous Malmquist bias can mimic a bulk flow (Landy & Szalay 1992). Without some *a priori* model of the density distribution $n(d)$ of the galaxies, it is difficult to remove the effect of inhomogeneous Malmquist bias from the observed data to reconstruct the true distribution.

Since the simulation shown in Figure 10 is a good representation of the observed spatial and velocity distribution functions, it provides a useful estimate of the effects of the inhomogeneous Malmquist bias on the observed distribution function. For the "galaxies" in the simulation, since we know both distances and velocities, we have prior knowledge of the density distribution $n(d)$.

We simulate "observations" by subjecting each galaxy in the simulation to a correction for (a) a homogeneous Malmquist bias with $\Delta = 0.36$ mag as described in §2.1 and used throughout this paper, and (b) an inhomogeneous Malmquist bias following Landy and Szalay (1992). For the latter, we used a density distribution $n(d) \propto d^{2.5}$, as determined directly from the simulation. In each case, as shown by Landy and Szalay (1992), the resultant histogram is not centered on zero, showing that the correction has an effect similar to a bulk flow in the sample. We remove this effect by subtracting the mean peculiar velocity in five distance bins to center the histogram on zero (equivalent to a bulk flow correction).

These would therefore be the "observed" histograms if we had not corrected for the homogeneous Malmquist bias and inhomogenous Malmquist bias respectively.

Kyoto Simulation: Model S
Inhomogenous Malmquist Correction

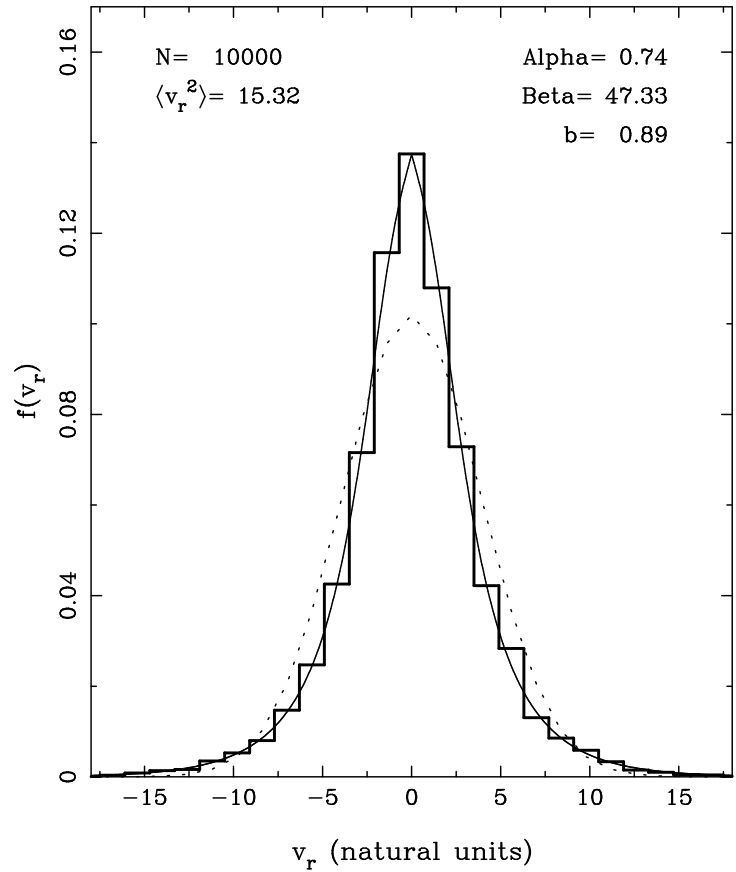

**Figure 11:** The contribution of corrections for inhomogeneous Malmquist bias to an observed sample. The 10000 particle Kyoto simulation (as used in Fig 10) is "observed" here with distances being perturbed by an inhomogenous Malmquist bias corresponding to $\Delta = 0.33$ and $n(d) \sim d^{2.5}$ as obtained from the simulation itself. We then subtracted the mean peculiar velocity in five distance bins to center the histogram on zero. Comparison of the resultant histogram with Figure 10 shows that inhomogenous Malmquist bias corrections cannot make substantial differences to the profile of the peculiar velocity distribution.

These produce almost identical distributions: in Figure 11 we show only the case of the inhomogeneous Malmquist bias. The histogram shows a slight displacement (about half a bin), but the best fit has hardly changed, and the difference from the Maxwell-Boltzmann distribution remains.